\documentclass[11pt,twoside]{article}
\usepackage{hepro5}
\usepackage{graphicx}

\usepackage[T1]{fontenc} % Computer Modern (CM) fonts
\usepackage{lmodern}

\usepackage{latexsym}
\usepackage{verbatim}
\usepackage{multirow}
\begin{document}

\vskip 1.0cm
\markboth{A.~Janiuk et al.}{Variability and winds in micro-QSOs}
\pagestyle{myheadings}

\vspace*{0.5cm}
\title{Non-linear variability in microquasars in relation with the winds from their accretion disks}

\author{A.~Janiuk$^1$, M. Grzedzielski$^1$, P. Sukova$^1$, F. Capitanio$^2$, S. Bianchi$^3$, W. Kowalski$^1$}
\affil{$^1$Center for Theoretical Physics,
Polish Academy of Sciences, Al. Lotnikow 32/46, 02-668 Warsaw, Poland\\
$^2$INAF-Instituto di Astrofisica e Planetologia Spaziali, via del Fosso del Cavaliere 100, 00133 Rome, Italy\\
$^3$Dipartimento di Matematica e Fisica, Universit\`a degli Studi Roma Tre, via della Vasca Navale 84, 00146 Roma, Italy}

\begin{abstract}

The microquasar IGR J17091-3624, which is the recently discovered analogue of the well known source GRS 1915+105, exhibits quasi-periodic outbursts, with a period of 5-70 seconds, and regular amplitudes, referred to as ``heartbeat state''. We argue that these states are plausibly explained by accretion disk instability, driven by the dominant radiation pressure. Using our GLobal Accretion DIsk Simulation hydrodynamical code, we model these outbursts quantitatively. We also find a correlation between the presence of massive outflows launched from the accretion disk and the stabilization of its oscillations. We verify the theoretical predictions with the available timing and spectral observations.
Furthermore, we postulate that the underlying non-linear differential equations that govern the evolution of an accretion disk are responsible for the variability pattern of several other microquasars, including XTE J1550-564, GX 339-4, and GRO J1655-40. This is based on the signatures of deterministic chaos in the observed lightcurves of these sources, which we found using the recurrence analysis method. We discuss these results in the frame of the accretion disk instability model.

\end{abstract}

\section{Introduction}

In black hole 
accretion disks, two main types of thermal-viscous instabilities may arise:
(i) the radiation pressure instability and (ii) the partial hydrogen ionization instability.
They can lead to
(i) the short term limit-cycle oscillations in black hole X-ray binaries (from tens to hundreds of seconds), to the intermittent activity of quasars (timescales from tens to thousands of years), or (ii) to the X-ray novae eruptions (timescales from months to years), and to the long-term activity cycles in AGN (timescales about millions of years), respectively. Both these types of instabilities are known in theoretical astrophysics for about 40 years (Lightman \& Eardley 1974, Smak 1984). In both cases, the variability and cyclic outbursts of the accretion disk are governed by the nonlinear equations of hydrodynamics.
However, as long as for the partial hydrogen ionization instability the
observations of Dwarf Novae systems 
have already confirmed the theoretical predictions
in many ways, the radiation pressure instability occurrence has been the subject of extensive debate. In particular, the only certain candidate for this instability being in action in the case of Galactic microquasars 
is the source GRS 1915+105. Nevertheless, there are many other systems, in
which the mean accretion rate is high enough (i.e., more than about 10\% of Eddington rate) to allow for the occurrence of the hot, radiation pressure dominated regions in the innermost accretion flow. Consequently, the non-linear variability of the observed flux at short timescales should be observed also in 
other sources.

It has been shown that the accretion disk can be stabilized partially or completely 
against these instabilities by  several mechanisms.
First, the very strong jet or wind outflow can reduce the amplitudes of
oscillations and outbursts (Janiuk \& Czerny 2011).
Also, a modified heating prescription, in which the viscous stress tensor scales not with the total pressure, but with a geometrical mean of the gas and total pressures, is an obvious stabilizer (Czerny et al. 2009).
In eccentric binaries, the influence of a companion star may lead to the decrease of a mean accretion rate and hence stabilize the disk at some parts of the binary orbit (Kunert-Bajraszewska \& Janiuk 2011). 
Finally, the stochastic viscous fluctuations can be a viable mechanism to stabilize the cyclic variability of the accretion disk (Janiuk \& Misra 2012). 
Here we neglect the three latter possibilities, and we study the accretion disk, influenced by the radiation pressure, while some part of the locally dissipated energy flux is used to eject a wind from the disk's surface. We show that the competing roles of the instability and wind outflow lead to the long-term trend, in which the heartbeat oscillations of the X-ray luminosity are weaker, or cease completely, when the wind is strong.
%We confront and confirm our model with the available observations of the microquasar IGR J17091, with both the timing, and spectral analysis.
%Furthermore, we study the lightcurves of three other X-ray sources.
%In this case, we do not have the information about the wind properties, but we can tentatively hint for the signatures of non-linear hydrodynamics governed by the accretion disk instability in some states of these sources.

\section{Accretion disk model}
\label{instability}

We use our own numerical code GLADIS (GLobal Accretion Disk InStability) which simulates the behavior of an accretion disk  in the frame of
%around IGR-J17091. 
a 1,5-dimensional hydrodynamics (Janiuk et al. 2002). We assume a geometrically thin, Keplerian $\alpha$-disk  under the pseudo-Newtonian Paczynski-Wiita potential. The structure of the disk is solved according to the continuity and energy equations, in the latter the viscous heating is balanced by radiative and advective cooling. For a given mean accretion rate (i.e., the mass accreted from the companion star in an X-ray binary, which is a parameter in our model), some part of the disk is dominated by radiation pressure. Hence, the local accretion rate, density, temperature, and disk thickness, oscillate. This behavior is shown in quantitative way in Figure 1.

%\subsection{Wind from the accretion disk}
%\label{wind}

%The presence of wind is important to regulate the amplitudes of disk oscillations, to the level that reproduce the observed X-ray variability. 
%It has been also confirmed by the spectroscopic observations of the microquasar IGR J17091 (see below).

\begin{figure}  %%%%%%%%%%%%%%%%%%%%%%FIGURE 1 %%%%%%%%%%%%%%%%%%%%%%%
\begin{center}
\hspace{0.25cm}
\includegraphics[height=8.4cm]{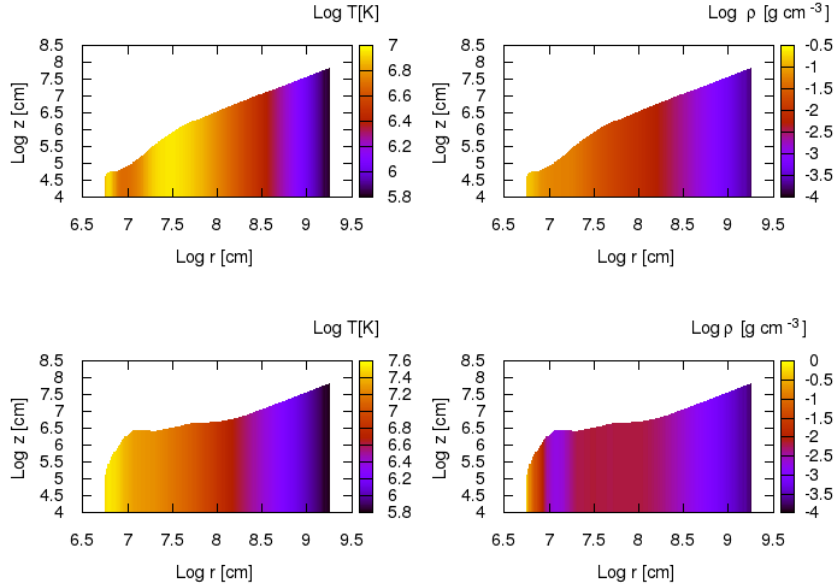}
\caption{Profiles of the temperature (left) and density (right) of the 
disk between the outbursts (top row) and at its outbursts maximum (bottom row). 
Colors present the quantities in a log-scale, in the r-z plane. 
Parameters of the simulation: mass of black hole 6 $M_{\odot}$, accretion rate 0.1 Eddington, viscosity $\alpha=0.1$. Also, a wind, with dimensionless strength coefficient $A=15$, was assumed.}
\label{fig1}
\end{center}
\end{figure}

\begin{figure}  %%%%%%%%%%%%%%%%%%%%%%FIGURE 2 %%%%%%%%%%%%%%%%%%%%%%%
\begin{center}
\hspace{0.25cm}
\includegraphics[height=5.0cm]{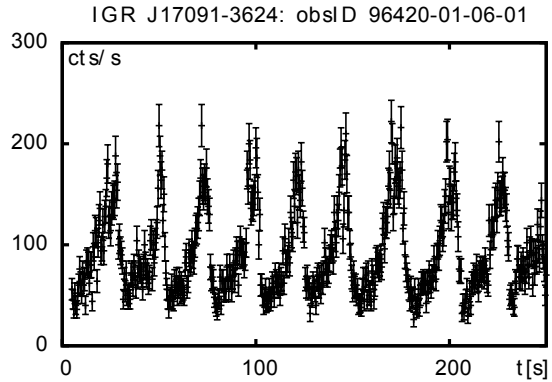}
\caption{X-ray lightcurve of the microquasar IGR J17091-3624 in its heartbeat state, taken from the PCA RXTE observations from April 4, 2011. Data in the 2-10 keV energy band, were extracted from the event files with time resolution of 0.5 s. 
%Obs ID is 96420-01-06-01.
}
%obr_IGR602_lc.eps
\label{fig2}
\end{center}
\end{figure}

Also, some fraction of energy that is locally dissipated and transferred vertically to the disk surface, is used to eject the wind, when the local disk luminosity instantaneously approaches the Eddington limit. In particular, we adopt a simple mathematical prescription for $f_{\rm out} = 1 - 1/(1+A \dot m^{2})$, with $\dot m (r,t) = L/L_{\rm Edd}$ (Nayakshin et al. 2000). We are then able to compute the mass loss rate from the disk, as equal to the ratio of the locally generated energy flux, to the energy change per particle, which is given by the virial energy. The total mass loss, $\dot M_{\rm w}$, is the one integrated over the disk surface, between some radius $R_{\rm min}$ and size of the disk $R_{\rm max}$ (see Janiuk et al. 2015, for further details).

\section{Observations}
\label{obs}

\subsection{Case study of IGR J17091-3624. Variability and wind}
\label{igr}
The outburst of the microquasar IGR J17091-3624 observed in 2011 was one of its brightest (Capitanio et al. 2012).
According to Altamirano et al (2011), in this source, in comparison to GRS 1915+105, two clear differences occur: 
(i) the time scales can be different (IGR J17091-3624 tends to be faster), and 
(ii) the average count rate (or flux) of the source can be much higher (factor 10-50) in GRS 1915+105.
If  the period of the oscillations is proportional to some power of the mass of the compact object 
%(see, e.g., Belloni et al. 1997; Frank et al. 2002), 
then the black hole in IGR J17091-3624 could be a factor of a few less massive than the $14 \pm  4. 4 M_{\odot}$  of GRS 1915+105. In our modeling, we assume the black hole mass to be 6 Solar masses.

In Figure 2, we show an X-ray lightcurve of this source, which is representative for the heartbeat oscillation. The amplitude of these flares is well modeled with our instability code, provided that we incorporate a moderate wind
with its 'strength' parameter on the order of $A=15$. The mass loss rate computed from such model is consistent with the upper limit, derived from the \textit{Chandra} 
observations (see Janiuk et al. 2015 for details).
In the non-heartbeat state, the wind is required by the model fitting in the \textit{Chandra} spectra, which allowed us to constrain the winds velocity $v$, its ionization parameter $\xi$ and filling factor $f$.
In Table 1, we summarize these results, together with the mass loss rate and the radial extension of the wind launching zone, derived from our numerical model.
In the non-heartbeat state, the wind strength parameter $A$ has to be much larger, so that it suppresses the oscillations of the disk completely.

\begin{table*} %%%%%%%%%%%%%%%%%%%%%%TABLE 1 %%%%%%%%%%%%%%%%%%%%%%%
%\begin{center}
\caption[]{Wind parameters. Observations vs. model in the 'heartbeat' (HB) and 'non-heartbeat' (NHB) states in the microquasar IGR J17091-3624. Two wind components, w1 and w2, were detected in the NHB \textit{Chandra} data (see Janiuk et al. 2015 for details).}
 \label{tw}
\begin{tabular}{|c|c||c|c|c|c||c|c|c|c|}
 \hline
\multicolumn{2}{|c||}{} &  \multicolumn{4}{|c||}{Observations} & \multicolumn{4}{|c|}{Model}
\\
 \hline
    & Wind  & $v$         & $\log \xi$      & $n$     & $f$    &   $\dot M_{\rm w}$     &
$R_{\rm min}$   & $R_{\rm max}$ & \\
State &  & [km s$^{-1}$] &       & [cm$^{-3}$] &     & [g s$^{-1}$] & $[R_{\rm g}]$ & [$R_{\rm g}$] & $A$ \\
 \hline
\multirow{2}{*} {NHB} & w1 & $9700 {\pm 800}$ & $3.4^{\pm 0.3}$ & $5.1 \cdot 10^{15}$ & 0.0015 & $2.7 \cdot 10^{17}$ & 950 & 4200 & 300
\\
%\hline
\cline{2-10}
               & w2 & $15700 {\pm 600}$& $3.8^{\pm 0.2}$ & $1.3 \cdot 10^{16}$ & 0.0037 & $4.2 \cdot 10^{17}$ & 380 & 4700 & 300 
\\
\hline

\multirow{2}{*} {HB} & w1 & -  & -  & $\le 5 \cdot 10^{14}$ & - & $2.5 \cdot 10^{16}$ & 950 & 4900 & 15
\\
%\hline
\cline{2-10}
            & w2 & -  & - & $\le 10^{15}$ & - & $3.9 \cdot 10^{16}$ & 380 & 5900 & 15
\\
\hline
\end{tabular}
%\end{center}
\end{table*}

\subsection{Other X-ray binaries. Non-linear variability}
\label{other}

\begin{figure}  %%%%%%%%%%%%%%%%%%%%%%FIGURE 3 %%%%%%%%%%%%%%%%%%%%%%%
\begin{center}
\hspace{0.25cm}
\includegraphics[height=5.0cm]{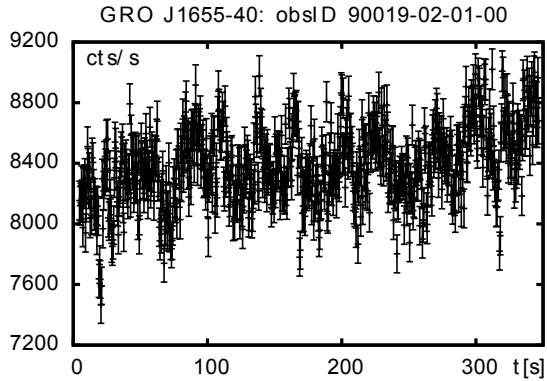}
\caption{X-ray lightcurve of the microquasar GRO J1655-40
in its outburst from 2005. 
%The obs ID is 90019-02-01-00, and 
The source's spectral state was classified as Soft-Intermediate, according to 
Miller et al. (2008). The data in the energy range 2-10 keV, from PCA/RXTE,
taken on March 13, 2005, 
were extracted in Std1 mode, with time resolution of 0.5 s.
%XTE J1550-564. Obs ID 30188-06-03-00}
}
\label{fig3}
\end{center}
\end{figure}

For the X-ray sources other than the two above-mentioned microquasars, no detailed analysis was performed before, with respect to their possible non-linear variability induced by the thermal-viscous instability in the accretion flow.
As proposed by Janiuk \& Czerny (2011), at least eight of the known black hole binaries have their accretion rates high enough for their disks to be unstable.
In the recent work by Sukova et al. (2016), we studied several of these sources.
We analyzed quantitatively their observed lightcurves, in order to find the traces of deterministic chaos type of behavior. If found, it indicates that the non-linear hydrodynamical process is acting behind the observed variability, and is governed by a finite set of differential equations. Therefore, 
the global behavior of the accretion flow, rather than some stochastic processes, is responsible for the shape of the observed lightcurves. We suggest that the disk instability is operating also in the accretion flow (possibly affected by the presence of a non-thermal corona above the disk), even if the regular 'heartbeat' flares are not directly visible in the lightcurves.

An exemplary lightcurve of  GRO J1655-40,
%XTE J1550-564, 
one of the sources which we studied in our sample, is shown in Figure 3. In this case, our method, based on the recurrence analysis, gave a significant result for deterministic chaos type of behavior.
%\subsection{Recurrence analysis}
%\label{chaos}
Basic object of the method is the recurrence matrix, defined as 
\begin{equation}
\mathbf{R}_{i,j}(\epsilon) = \Theta (\epsilon - \parallel \vec{x}_i - \vec{x}_j \parallel ), \qquad i,j = 1,...,N, \label{RP_def}
\end{equation}
where $\vec{x}_i = \vec{x}(t_i)$ are ($N$) points of the reconstructed trajectory, $\Theta$ is the Heaviside step function, and  $\epsilon$ is a threshold parameter. It can be visualized in the recurrence plot (for details, see Sukova et al. 2016).
%The long diagonal lines in such plot represent the situation, when the trajectory (reconstructed from the time-series by time delay technique) returns close to itself in two different times.
We compared the results between real and surrogate data (the latter have the same power spectra, but variability is stochastic). 
 The significance of chaos is defined as a weighted difference between the logarithm of the second order Renyi's entropy of the data series and its surrogates sample.
Our method was tested with simulated trajectories of complicated non-linear systems, i.e., motion of the test particle in the field of a black hole, given by Einstein equations (see article Sukova \& Janiuk, in these proceedings). Its chaotic orbit shows high significance of non-linear dynamics (Semerak \& Sukova 2012).
Highly significant results were obtained also for the microquasars XTE J550-564, and GX 339-4.
%We examined several observations of the other five microquasars. Aside from the well-studied binary GRS 1915+105, we found significant traces of non-linear dynamics also in three other sources (GX 339-4, XTE J1550-564 and GRO J1655-40). 
 The non-linear behavior of the lightcurve during some of the observations gives the evidence, that the accretion flow in these sources is governed by a low number of non-linear equations. A possible explanation is that the accretion disc is prone to the thermal-viscous instability and the induced limit-cycle oscillations.

%obr_IGR602_Lmax.eps
%\begin{figure}  %%%%%%%%%%%%%%%%%%%%%%FIGURE 4 %%%%%%%%%%%%%%%%%%%%%%%
%\begin{center}
%\hspace{0.25cm}
%\includegraphics[height=5.0cm]{hepro5_fig4.eps}
%\caption{Dependence of the longest diagonal line $L_{\rm max}$ given in points on the recurrence threshold $\epsilon$ for the observation of IGR J1709 presented in Figure 1 (red color). The ensemble of 100 surrogates is shown in grey color.
%!!!cyan: remove!!! 
%and the ensemble of shuffled surrogates are shown in cyan.
%}
%\label{fig4}
%\end{center}
%\end{figure}

\section{General picture and conclusions}
\label{discussion}

IGR J17091-3624 is another microquasar, after GRS 1915+105, that in some states exhibits the limit-cycle oscillations of its X-ray luminosity. 
These oscillations are plausibly explained by the intrinsic thermal-viscous instability of the accretion disk, induced by the radiation pressure.
The fast, ionized wind ejected from the accretion disk on the cost of a fraction of dissipated energy is a viable mechanism to completely stabilize the disk in other states, or to govern the moderate amplitude of the disk oscillations.

In other black hole X-ray binaries, i.e. GRO J1655-40, XTE J1550-564, and GX 339-4, the hints of a non-linear variability were also found, using the novel method, which adapts the recurrence analysis for the study of a deterministic chaos process.
%Possible further mechanism stabilizing the oscillations in other sources or different spectral states is a stochastic fluctuation of the viscosity (modeled as a Markov chain process; see Lyubarskii 1997; King et al. 2004; Janiuk \& Misra 2012).
We applied this analysis to observations of six black hole  X-ray binaries observed by the RXTE satellite, in their soft and intermediate states.
% We developed a method for distinguishing between stochastic, non-stochastic linear and non-linear processes using the comparison of the quantification of recurrence plots with the surrogate data. 
We also tested this method on the sample of observations of the microquasar IGR J17091-3624, in its various spectral states defined as in Pahari et al. (2014). Significant results for the ``heartbeat'' state were obtained, which confirmed our previous findings for this source.

We suggest a possible geometrical configuration of the accretion flow in IGR J17091-3624, and possibly other microquasars where an ionized wind is detected (e.g. GRO J1655-40). The radiation pressure dominated accretion disk is stabilized by
a quasi-static corona in its inner parts, and by an unbound wind ejected at its
outer radii. 
The wind (possibly with multiple components), 
may be partially collimated and be narrower in the part where the spectral features are produced.
%{\bf Describe flow Geometry and wind detection possibilities in other sources ???}

\acknowledgments This work was supported in part by the grant DEC-2012/05/E/ST9/03914 from the
Polish National Science Center.

\end{document}